\documentclass[pre,twocolumn,notitlepage,longbibliography,amsmath,amssymb,floats,superscriptaddress,nofootinbib,10pt]{revtex4-2}
\usepackage{graphicx,color}
\usepackage{amsfonts, amsbsy}
\usepackage{amssymb}
\usepackage{amsmath}
\usepackage{eqnarray}
\usepackage{bm}
\usepackage{blkarray}
\usepackage{mathtools}
\usepackage{psfrag}
\usepackage{caption}
\usepackage{subcaption}
\usepackage{multirow}
\usepackage{textcomp}
\usepackage{units}
\usepackage{lipsum}
\usepackage[title]{appendix}
\usepackage{soul}
\usepackage{titlesec}
\usepackage{times}
\usepackage{enumitem}
\usepackage{soul}
\usepackage[breaklinks,colorlinks = true,linkcolor = blue,urlcolor  = blue,citecolor = blue,anchorcolor = blue]{hyperref}
\usepackage{float}

\begin{document}
\title{Fracton-elasticity duality on curved manifolds}
\author{Lazaros Tsaloukidis}
\email{ltsalouk@pks.mpg.de}
\affiliation{Max Planck Institute for the Physics of Complex Systems, N\"othnitzer Stra{\ss}e 38, 01187 Dresden, Germany}
\affiliation{W\"urzburg-Dresden Cluster of Excellence ct.qmat, 01187 Dresden, Germany}
\author{Jos\'e J. Fern\'andez-Melgarejo}
\email{melgarejo@um.es}
\affiliation{Departamento de Electromagnetismo y Electr\'{o}nica, Universidad de Murcia, Campus de
Espinardo, 30100 Murcia, Spain}
\author{Javier Molina-Vilaplana}
\email{javi.molina@upct.es}
\affiliation{Departamento de Autom\'{a}tica. Universidad Polit\'{e}cnica de Cartagena, Calle Dr. Fleming, S/N, 30202 Cartagena, Spain}
\author{Piotr Sur\'{o}wka}
\email{piotr.surowka@pwr.edu.pl}
\affiliation{Institute of Theoretical Physics, Wroc\l{}aw University of Science and Technology, 50-370, Wroc\l{}aw, Poland}
\date{\today}

\begin{abstract}
The mechanical properties of crystals on curved substrates mix elastic, geometric and topological degrees of freedom. In order to elucidate the properties of such crystals we formulate the low-energy effective action that combines metric degrees of freedom with displacement fields and defects. We propose dualities for elasticity coupled to curved geometry formulated in terms of tensor gauge theories. We show that the metric degrees of freedom, evolving akin to linearized gravity are mapped to tensors with three indices. When coupled to crystals these degrees of freedom become gapped and, in the presence of dislocations and disclinations, multivalued. The elastic degrees of freedom remain gapless and mapped to symmetric gauge fields with two indices. In analogy with elasticity on flat space formulation we assume that the trace of the total quadrupole moment is conserved. In the dual formulation, topological defects, which act as sources for the gauge fields, are fractons or excitations with restricted mobility. This leads to generalized glide constraints that restrict both displacement and gravitational defects.
 \end{abstract}

\maketitle

\section{Introduction} Two-dimensional crystals on curved substrates emerge in a variety of systems including colloids \cite{fernandez-nieves_crystals_2016}, metamaterials \cite{bertoldi_flexible_2017} and biomatter \cite{lecuit_cell_2007}. In order to elucidate mechanical properties of such systems one requires a consistent framework of crystallography on curved manifolds. The crystallization of atoms on non-flat topographies naturally leads to the emergence of topological defects. This follows from the fact that the equilibrium structure of crystals is encoded in the arrangement of disclinations and dislocations. The goal of this paper is to formulate the low-energy effective description of curved crystalline fields and investigate the dynamics of defects. In order to achieve this goal we propose a duality between elasticity and gauge theories that incorporates both the metric degrees of freedom and elastic displacements. 

\begin{figure*}[th!]
        \centering
\includegraphics[width=0.9\textwidth]{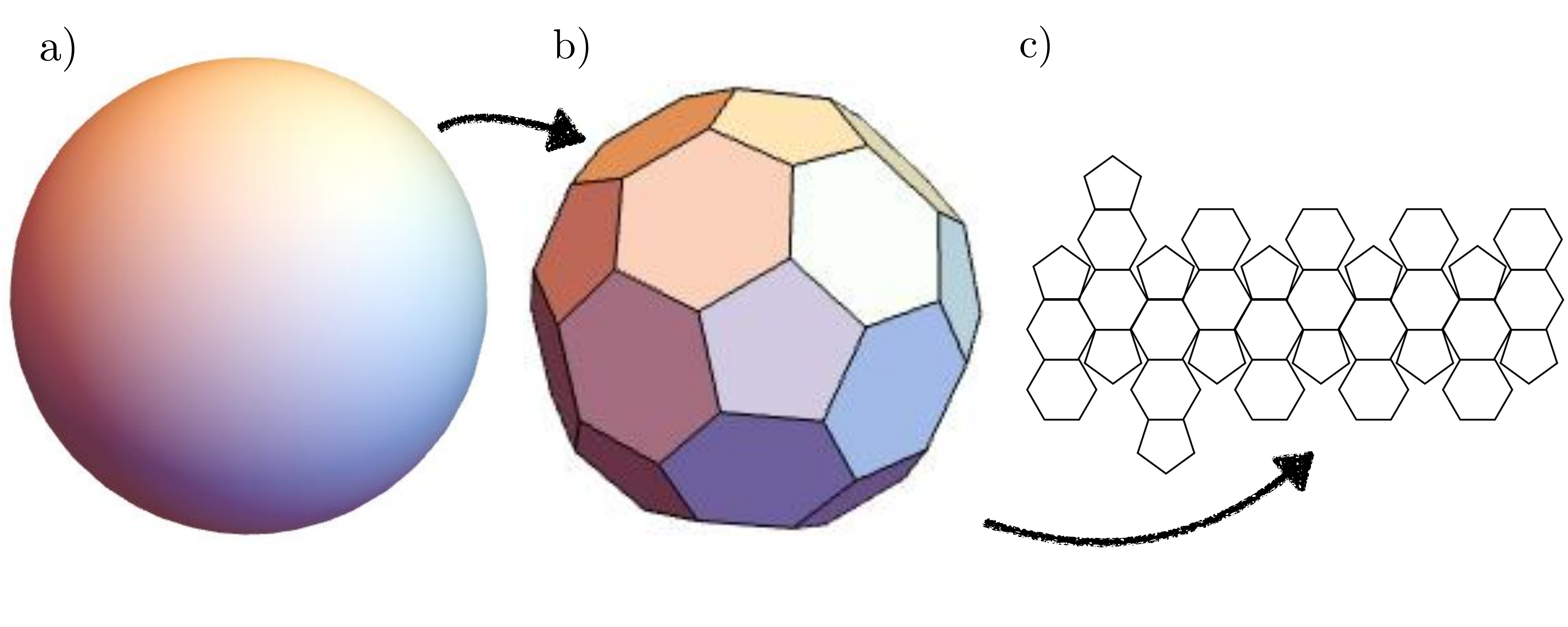}
        \caption{a) As an example of a crystal on a curved manifold we start with a sphere. b) We allow atoms to break the rotation symmetry spontaneously by creating a crystal on its surface. c) Curvature is visible by the necessity of including pentagons in a hexagonal lattice, which unfolded on a plane shows angle deficiency.}
        \label{fig:sphere}
    \end{figure*}

The flat-space theory of elasticity serves as a description of deformed solid bodies. In the traditional formulation the deformation is parameterized by a displacement field that captures the departure from the equilibrium configuration. The microscopic intuition behind the displacement field is given in terms of fluctuations of atoms around the equilibrium arrangement of their position on the lattice. The crystalline lattice results from a spontaneous symmetry breaking of translations in space. During crystal growth atoms can migrate from their lattice
positions and a mismatch in crystal planes can occur. This leads to the formation of defects. Vacancies in the lattice structure are point defects that do not cause additional stresses. On the other hand line or plane irregularities within a crystal create non-local stresses. We call such quantities topological. These stresses cannot be removed without cutting the crystal and global rearrangements of the lattice structure. In two dimensions topological defects are represented as points or singularities of the elastic fields. In most materials the defects are created because of impurities contained in the sample. However, in two dimensions, crystallization on a curved substrate provides another mechanism, that necessarily leads to the formation of defects (See Fig. \eqref{fig:sphere}). This conforms to the usual lore that topological defects are sources of curvature and torsion. Therefore a consistent description of a crystal on a curved substrate should account for geometry, defects and fluctuations of atoms forming the crystal \cite{kleinert_gauge_1989}.

Our construction extends modern developments in the theory of elasticity, namely fracton-elasticity dualities used to reformulate elasticity as gauge theories \cite{kleinert_duality_1982,kleinert_double_1983,beekman_dual_2017-1,beekman_dual_2017,pretko_fracton-elasticity_2018,pretko_crystal--fracton_2019,radzihovsky_fractons_2020,radzihovsky_quantum_2020,gromov_duality_2020,surowka_dual_2021,hirono_effective_2022,caddeo_emergent_2022} and crystal gravity employed to marry the dynamics of metric degrees of freedom with spontaneous symmetry breaking of crystals \cite{zaanen_crystal_2022}. We introduce topological defects as singularities that lead to multivalued fields both in elasticity and gravity. This extends earlier developments in this direction \cite{kleinert_multivalued_2008}. Our main results are the following: 1) elasticity on curved backgrounds admits the dual formulation in terms of tensor gauge fields, 2) disclinations and dislocations in symmetric elasticity on curved manifolds are fractons and partial fractons respectively, exhibiting mobility restrictions when certain constraints regarding total dipole and quadrupole moment conservation are imposed, in analogy with flat space. As a byproduct of our analysis we show that linearized gravity can be formulated as a gauge theory.

Our theory opens up many directions in elasticity in curved substrates. First of all it gives a consistent low-energy field theory approach to the problem. This is in contrast to previous developments that are rather phenomenological and do not fall in the effective field theory paradigm \cite{efrati_elastic_2009,moshe_elastic_2015,li_elasticity_2019}. In addition our work avoids many difficulties with the physical interpretation, in terms of spontaneous symmetry breaking, of the proposals in the mathematical literature \cite{ciarlet} or in general relativity \cite{Carter_Quintana_foundations_1972,beig_relativistic_2003}, which aim at a direct covariant formulation of elasticity. We can consistently study the dynamics of elastic fields and defects on curved backgrounds and provide a clear physical interpretation of the resulting dynamics. The formalism also allows for extensions to account for new degrees of freedom both in passive and active solids. It complements developments in soft matter physics, that focus on the geometric properties of crystal shells (see e.g. \cite{kamien_geometry_2002,vitelli_crystallography_2006,bowick_two-dimensional_2009}).

\section{Dual formulation of linearized metric fluctuations} As a first step before considering crystalline structures with curvature, we start with a membrane described by a linearized theory of metric fluctuations in $2+1$ dimensions
\begin{equation}
S=\frac{1}{2}\int d^2xdt \left[(\partial_th_{ij})^2-\kappa(\partial_kh_{ij})^2\right], \label{eq:lingrav}
\end{equation}
where $h_{ij}$ refers to a small metric fluctuation around a fixed background and $\kappa$ is a constant parameter. This is the Fierz-Pauli action in the helical projection when the time components are omitted. In general $\kappa$ could be a constant tensor, however, this renders the expressions more cumbersome. Since it does not affect the dualization procedure, we only choose to work with the most asymmetric term comprising the theory. Notice that similar models have been considered in the context of polymerized membranes \cite{paczuski_landau_1988} (for a review see \cite{nelson_defects_2002}) or linearized gravity \cite{Fierz_Pauli_1939} (for a review see \cite{hamber_quantum_2008}).

In our case, Eq. \eqref{eq:lingrav} can be viewed as a toy model to set the stage for elasticity on a curved background and illustrate the steps that are necessary to construct the duality for a theory that contains a metric. We ask the question if this theory has a dual formulation as a gauge theory. In fact such a question in the context of gravity has a long history with different approaches that range from rewriting gravity in terms of gauge variables \cite{witten_2_1988} to developments based on gauging the algebraic structure of gravity \cite{utiyama_invariant_1956,kibble_lorentz_1961,sciama,ivanov_gauge_1982,grignani_gravity_1992}. In this work, having in mind applications to elasticity, we perform a Hubbard-Stratonovich transformation on Eq. \eqref{eq:lingrav}, which is at the core of elasticity/gauge theory dualities. For given sets of conjugate tensor fields $\phi $ and $\psi$ we can compactly write it as 
\begin{equation}
\exp \left[\frac{1}{2} {\psi M \psi} \right]=\frac{1}{\mathcal{N}}   \int D \phi \exp \left[{\frac{-1}{2}  \phi M^{-1} \phi + \psi \phi} \right], \,
\end{equation}
 where $\mathcal{N}$ is a normalization factor. Here we take the conjugate fields to be $L_{ij}=\partial_th_{ij}$, $K_{kij}=\kappa \partial_kh_{ij}$. It is important to note that no gauge choice is imposed before the dualization, which is standard in the previous approaches to crystallography on curved substrates \cite{vitelli_crystallography_2006}. Picking a gauge may introduce higher-order terms in the action, which obstructs the Hubbard-Stratonovich transformation. So far we have assumed that the metric is smooth. We can also introduce singularities by the following splitting $h_{ij}=\Bar{h}_{ij}+h^{(s)}_{ij}$, with $(s)$ denoting the singular part. The dual action is
\begin{multline}
S_{\text{HS}}=\frac{1}{2}\int d^2xdt\left[\frac{1}{\kappa}(K_{kij})^2-(L_{ij})^2-\right.
\\ 
-2K_{kij}\partial_kh^{(s)}_{ij}+2L_{ij}\partial_th^{(s)}_{ij}\Bigl].
\label{GHS}
\end{multline}
In deriving Eq. \eqref{GHS}, we have integrated out the smooth part, leading to the conservation law
\begin{equation} \label{eq:contthreetens}
\partial_tL_{ij}-\partial_kK_{kij}=0 .   
\end{equation}
This equation is just a dual representation of the wave equation for the metric distortions. Thus it captures the dynamics of the substrate. The defects of the theory are points in space where partial derivatives do not commute. Their density form, as well as the current associated with them, are given by:
\begin{subequations}
\begin{align}\label{eq:rhotilde}
\Tilde{\rho}_{ij}=&\ \epsilon_{ik}\epsilon_{jl}\epsilon_{mn}\partial_n\partial_mh^{(s)}_{kl} , 
\\[8pt]
\label{eq:currenttilde}
\Tilde{J}_{mij}=&\ \epsilon_{ik}\epsilon_{jl}(\partial_m\partial_t-\partial_t\partial_m)h_{kl}^{(s)} .  
\end{align}
\end{subequations}
The defect density is described by a tensor density. Eq. \eqref{eq:contthreetens} is a continuity equation that, as it will be shown below, can be resolved in terms of gauge fields $\Tilde{{A}}_{lij}$ and $\Tilde{\phi}_{ij}$. We present the details in a combined theory, in which the coupling of $h_{ij}$ to elasticity has been introduced. \\
\section{Elasticity on a curved background} Crystallization occurs as a spontaneous symmetry breaking of the translation symmetry. Such a process does not require the underlying manifold to be flat. However, if a lattice forms on a curved substrate the curvature is encoded in topological defects that cannot be avoided. It is important to note that defects are singularities in the macroscopic fields, which in this case correspond to displacements or lattice vibrations and the metric. As a result, contrary to flat lattices, the equilibrium configuration contains defects. If new defects are created in the equilibrium configuration, they affect both metric and displacement degrees of freedom. Before we study the dynamics of defects we focus on a region of space without defects. Such a configuration can be understood similarly to the spontaneous symmetry breaking present in superconductors. Here, the strain field, plays the role of the order parameter for the metric \cite{zaanen_crystal_2022}. The low-energy theory that describes both the rigidity of the solid plate and the linearized fluctuations of the metric is given by
\begin{multline}
S=\frac{1}{2}\int d^2xdt[\rho_d(\partial_tu_i)^2+(\partial_th_{ij})^2-\\ \nonumber
-(u_{ij}+h_{ij})C_{ijkl}(u_{kl}+h_{kl})-\kappa(\partial_kh_{ij})^2].
\end{multline}
In the above, $\rho_d$ is the material density, $u_i$ is the displacement field, $u_{ij}=\frac{1}{2}(\partial_iu_j+\partial_ju_i)$ the strain field and $C_{ijkl}$ the usual 4-rank elasticity tensor whose exact form is dictated by the symmetries present in the crystal. For simplicity we expand around flat space, however, the procedure can be generalized as an expansion around arbitrarily curved substrates. An important aspect of this theory is that the metric degrees of freedom, or "gravitons" become gapped, while the fluctuations of atoms remain gapless. Spontaneous symmetry breaking of elasticity requires massive St\"{u}ckelberg fields as already anticipated in \cite{gromov_duality_2020,pena-benitez_fractons_2023}. Here the tensor of elastic coefficients plays the role of mass parameters. We note that a generic displacement fluctuation does not belong to the original manifold and has to be accompanied by a fluctuation of geometry. 

We now extend the theory to include defects in the picture. In order to do this consistently, we need to account both for singularities of the displacement field and the metric. An attempt in this direction was already done in \cite{kleinert_multivalued_2008}, where a purely geometric theory was constructed. However, we find it convenient to keep both the displacement fields and the metric. Thus we proceed by splitting down the displacement field into its singular and smooth phonon pieces. Analogously the metric is also divided into singular and smooth parts. The reason to keep singularities in the two fields can be understood as follows. Imagine that we have singular configurations of the displacement field, then 
\begin{equation}
u_{ij}+h_{ij} \rightarrow u_{ij} + h_{ij} +\frac{1}{2}(\partial_iu_j^{(s)}+\partial_ju_i^{(s)}).
\end{equation}
However now, we can also view this as a singular contribution to the metric
\begin{equation}
h_{ij}^{(s)} = \frac{1}{2}(\partial_iu_j^{(s)}+\partial_ju_i^{(s)}).
\end{equation} Therefore, without imposing any additional conditions, we can have singularities contributing both to the strain and the metric. 

We dualize the resulting effective action by means of the Hubbard-Stratonovich transformation to obtain
\begin{align} \label{eq:HSgrav}
&S_{\text{HS}}=\frac{1}{2}\int d^2xdt\left[T_{ij}C_{ijkl}^{-1}T_{kl}-\frac{1}{\rho_d}(\pi_i)^2+\frac{1}{\kappa}(K_{kij})^2\right.\nonumber \\ 
&\left.-(L_{ij})^2+2\pi_i\partial_tu_i^{(s)}+2L_{ij}\partial_th_{ij}^{(s)}-2w_{ij}^{(s)}T_{ij}\right.\\ &\left.-2\partial_kh_{ij}^{(s)}K_{kij}\right] ,   \nonumber 
\end{align}
where, $T_{ij}$ is the generalized stress-tensor, $\pi_i$ is the momentum density and $K_{kij}$ and $L_{ij}$ are their linearized gravity counterparts and $w_{ij}^{(s)}=u_{ij}^{(s)}+h_{ij}^{(s)}$. The dual formulation of this theory parallels Cosserat elasticity \cite{gromov_duality_2020}, where gapless displacements are supplemented with gapped bond angles. By performing integration by parts on the last four terms of Eq. \eqref{eq:HSgrav}, we obtain the two dynamical laws governing the system
\begin{align} \label{HookeLaw}
\partial_t \pi_i-\partial_jT_{ij}=&\ 0 ,
\\[8pt]
\label{GravityNewton}
\partial_tL_{ij}-\partial_kK_{kij}=-&\ T_{ij}.
\end{align}
Equation \eqref{HookeLaw} above is the well-known Hooke's law with the stress tensor containing contributions from the bending of the plate besides stretching and shearing. In symmetric elasticity it is resolved by introducing a first set of gauge fields ($\phi$, $A_{ij}$)
\begin{equation}
\pi_i=-\epsilon_{ij}\epsilon_{kl}\partial_kA_{lj}, \quad T_{ij}=-\epsilon_{ik}\epsilon_{jl}(\partial_tA_{kl}+\partial_k\partial_l\phi) .
\label{gaugefields1}
\end{equation}
Now, as it is commonly done in the context of elasticity/gauge duality, it is convenient to define the generalized electric and magnetic fields as
\begin{equation}
B_i=\epsilon_{ij}\pi_j, \qquad E_{ij}=\epsilon_{ik}\epsilon_{jl}T_{kl}.
\end{equation}
In this way Eq. \eqref{HookeLaw} takes the Faraday equation form. The second equation at hand, Eq. \eqref{GravityNewton}, captures the dynamics of defects from the metric fluctuations with the stress-tensor acting as a source. Here we introduce a second set of gauge fields $(\tilde \phi_{ij}, \ \tilde A_{ijk})$, first by plugging the $T_{ij}$ from Eq. \eqref{gaugefields1}, to get:
\begin{align}
&\ \Tilde{B}_{ij}=L_{ij}-\epsilon_{ik}\epsilon_{jl}A_{kl}=\epsilon_{ik}\epsilon_{jl}\partial_m\Tilde{A}_{mkl},
\label{TildeMagneto}
\\[8pt]
\Tilde{E}_{mij}=&\ K_{mij}+\epsilon_{im}\epsilon_{jl}\partial_l\phi=\epsilon_{ik}\epsilon_{jl}(\partial_t\Tilde{A}_{mkl}+\epsilon_{mn}\partial_n\Tilde{\phi}_{kl}).
\label{TildeElectro}
\end{align}
Note that while the rank-2 magnetic field tensor $\Tilde{B}_{ij}$ is symmetric under the permutation ($i\leftrightarrow j$), the rank-3 electric field tensor $\Tilde{E}_{kij}$ is not. In total the theory includes four gauge fields, closely similar to the Cosserat theory of elasticity.

Eq. \eqref{GravityNewton} means that $L_{ij}$ is not a conserved quantity, however, both $L_{ij}$ and $K_{kij}$ are observables of the theory and they should be invariant under a specific class of gauge transformations that leaves both of these quantities unchanged, i.e. $\delta L_{ij}=0$ and $\delta K_{kij}=0$. These gauge transformations can be constructed explicitly
\begin{align}
& \hspace{0.5cm} \delta\phi =\partial_t f ,\hspace{1.5cm} \delta A_{ij}=-\partial_i\partial_jf ,\\
\delta \Tilde{\phi}_{ij}&=-\partial_t\Tilde{f}_{ij},\, \qquad  \delta \Tilde{A}_{mij}=\delta_{m\left(i\right.}\partial_{\left.j\right)}f+\epsilon_{mn}\partial_n\Tilde{f}_{ij} \, ,
\end{align}
where $f(\vec{x},t)$ is any space-time scalar function and $\Tilde{f}_{ij}(\vec{x},t)$ is another space-time dependent tensor that is symmetric under exchange $i\leftrightarrow j$.
By using equations \eqref{TildeMagneto} and \eqref{TildeElectro} now, we can formulate the dual Lagrangian of the theory as function of the gauge fields
\begin{align} \label{duallagrangian}
&\mathcal{L}_{\text{dual}}=\frac{1}{2}\left[\Tilde{C}^{-1}_{mnab}E_{mn}E_{ab}-\frac{(B_i)^2}{\rho_d}+\frac{1}{\kappa}(\Tilde{E}_{kij}-\epsilon_{ik}\epsilon_{jl}\partial_l\phi)^2\right. \nonumber\\
&\left.-(\Tilde{B}_{ij}+\epsilon_{ik}\epsilon_{jl}A_{kl})^2\right]+\rho\phi-A_{ij}J_{ij}+\Tilde{\rho}_{ij}\Tilde{\phi}_{ij}-\Tilde{A}_{kij}\Tilde{J}_{kij},
\end{align}    
where we have defined $\Tilde{C}^{-1}_{mnab}=\epsilon_{mi}\epsilon_{nj}\epsilon_{ak}\epsilon_{bl}C^{-1}_{ijkl}$ for the rotated inverse elasticity tensor and we have also introduced the sources of defects, $\rho$, $\Tilde{\rho}_{ij}$ and currents $J_{ij}$, $\Tilde{J}_{kij}$ in our theory. The first two terms and their respective sources in the equation above, are the familiar theory of fracton-elasticity duality \cite{pretko_fracton-elasticity_2018}. From Eq. \eqref{duallagrangian} the continuity equations read
\begin{align}
\partial_t\rho-\partial_i\partial_jJ_{ij}-\partial_j\Tilde{J}_{iij}=0\, ,
\label{phononcontinuity}
\\[8pt]
 \qquad \partial_t\Tilde{\rho}_{ij}-\epsilon_{nm}\partial_n\Tilde{J}_{mij}=0 ,
\label{gravitoncontinuity}
\end{align}
with the sources and currents
\begin{subequations}
\begin{align}
\rho=&\ \epsilon_{ik}\epsilon_{jl}\partial_i\partial_j u_{kl}^{(s)}
\label{mixeddefect},
\\[8pt]
J_{ij}=&\ \epsilon_{ik}\epsilon_{jl}(\partial_t\partial_k-\partial_k\partial_t)u_l^{(s)},
\end{align}
\end{subequations}
and $\Tilde{\rho}_{ij}$ and $\Tilde{J}_{kij}$ given by Eqns. \eqref{eq:rhotilde} and \eqref{eq:currenttilde} respectively. In order now to study the fractonic behavior of the system, we begin by calculating the canonical momenta conjugate to the tensor fields $\Pi_{ij}=\left(\frac{\partial \mathcal{L}}{\partial \Dot{A}_{ij}}\right)$ and $\Tilde{\Pi}_{ijk}=\left(\frac{\partial \mathcal{L}}{\partial \Dot{\Tilde{A}}_{ijk}}\right)$. 
The Hamiltonian density then reads
\begin{equation}
\mathcal{H}=\Pi_{ij}\Dot{A}_{ij}+\Tilde{\Pi}_{ijk}\Dot{\Tilde{A}}_{ijk}-\mathcal{L}.
\label{Hamilton}
\end{equation}
Requiring gauge invariance of Eq. \eqref{Hamilton}, we arrive at
\begin{equation}
\partial_i\partial_j\Pi_{ij}+\partial_j\Tilde{\Pi}_{iij}=-\rho,
\end{equation}
\begin{equation}
\epsilon_{km}\partial_m\Tilde{\Pi}_{kij}=-\Tilde{\rho}_{ij}.
\end{equation}
To demonstrate that disclinations behave similarly to fractons, we evaluate the contributions from the two types of charges present to the system's overall dipole moment. We then proceed to derive the total dipole moment
\begin{align}
&P_a+\Tilde{P}_a=\int \left(\rho  x_a +\epsilon_{ia}\Tilde{\rho}_{ij}x_jd^2x\right)d^2x=\hspace{0.5cm}\nonumber\\
& \hspace{3.5cm}=\frac{1}{\kappa}\int K_{aii}d^2x+\text{const}.
\label{totaldipoleconserv}
\end{align}
The relation above extends the one originally formulated by Pretko \cite{pretko_fracton-elasticity_2018,pretko_crystal--fracton_2019} for a flat background. As in that case, the second constant term on the right-hand side arises from the integrals of the boundary terms, which are unaffected by local operations within the bulk of the region. In this context, we observe that the total dipole moment for disclination defects is not universally conserved. Intriguingly, it is connected to the behavior of specific components of the metric and only becomes a conserved quantity when $K_{aii}=0$. This is comparable to the situation in Cosserat elasticity, where two separate dipole moments are linked to two different kinds of defects, as indicated in \cite{gromov_duality_2020}.
This result indicates that disclinations can manifest fractonic properties. To this end, we follow \cite{pretko_fracton-elasticity_2018} and express the integrand in Eq. \eqref{totaldipoleconserv} in terms of elastic variables as,
\begin{equation}
\frac{1}{\kappa}K_{aii}=\partial_ah_{ii}=\partial_a(\Bar{h}_{ii}+h^{(s)}_{ii})=0   \, . 
\end{equation}
The smooth part represents volume changes attributed to the regular deformations in the metric. These, as noted in \cite{pretko_fracton-elasticity_2018} regarding creation of vacancies in the crystal can be set to zero i.e. $\partial_a\Bar{h}_{ii}=0$. Consequently, the singular component now signifies volume alterations that are contingent on the disclination defects present in the metric. Should that term be zero, as expressed by
\begin{equation}
    \partial_ah_{ii}^{(s)}=0\, .
    \label{metricconstraint}
\end{equation} 
then disclinations will indeed behave like fractons, as the total dipole moment of the system becomes conserved. An immediate result of all the above is that the trace of the metric defects density $\Tilde{\rho}_{ii}$ is now equal to zero (by use of Eq. \eqref{gravitoncontinuity} we also get $\epsilon_{nm}\partial_nJ_{mii}=0$).\\
\section{Glide constraint on curved manifold} A comparison with the theory on flat background gives us a physical intuition about the underpinning mobility restrictions mechanism. There, a physical requirement that the solid does not contain vacancies and interstitials leads to the so-called glide constraint, which states that the trace $J_{ii}$ vanishes \cite{cvetkovic_topological_2006}. In our work this is no longer the only requirement. Calculating the quadrupole moment from both contributions results in
\begin{align}
Q_{ii}+\Tilde{Q}_{ii}&=\int\left(\rho x_i^2+\epsilon_{im}\Tilde{\rho}_{ij}x_jx_m\right)d^2x=\nonumber\\
& =\int \left[2\Tilde{C}_{iikl}^{-1}T_{kl}+\frac{1}{\kappa}K_{aii}x_a\right]d^2x+\text{const}.
\end{align}   
In the relation above, the first term on the right hand side is the familiar result of Pretko \cite{pretko_crystal--fracton_2019} regarding volume changes connected to the presence of vacancies or interstitials in the crystal (expressed in elastic variables instead of their electromagnetic
counterparts). Since the stress tensor $T_{ij}$ is dual to the generalized strain tensor $w_{ij}$, this term now captures contributions to the volume change coming from the trace of the metric too. Should that term vanish along with Eq. \eqref{metricconstraint}, imposing dipole moment conservation, then the trace of the total quadrupole moment of the system is also a conserved quantity. Dislocations then become lineons, i.e. only able to move along their respective Burgers vector. The new modified constraint on a curved background can be formulated as two distinct sub-cases \begin{equation}
\text{1st case}: \quad u_{ii}^{(s)}=h_{ii}^{(s)}=0, \qquad J_{ii}=0,\, \hspace{0.5cm} \Tilde{J}_{mii}=0 ,
\end{equation}
\begin{equation}
\text{2nd case}: \quad u_{ii}^{(s)}=-h_{ii}^{(s)} ,\quad \Tilde{J}_{mii}=-(\partial_mJ_{ii}+\partial_m\partial_i\pi_i^{(s)})  ,
\end{equation}
where $\pi_i^{(s)}$ acts as the current for vacancies-interstitials of the displacement field. The first case is simply the generalized glide constraint in the curved background. The traces of both kinds of currents represent the defects present in the theory. In the second case the presence of any vacancies or interstitials originating from the displacement field are canceled out by their counterparts coming from the metric tensor. This cancellation leads to a relation between the two kinds of currents in the theory.

We conclude that disclocations can retain their fractonic behavior on curved manifolds provided that their motion perpendicular to the Burgers vector is not induced either by the presence of vacancies or interstitials or certain configurations of the strain tensor and the metric.

\section{Discussion}

In this work we have shown that elasticity on curved manifolds can be formulated in terms of tensor gauge variables. Elastic defects remain fractonic and exhibit mobility constraints, provided that one of the two generalized glide-constraints is satisfied. In accord with previous expectations we show that the duality does not map elasticity to tensor gauge theories coupled to curvature, which would break the gauge invariance \cite{slagle_symmetric_2019,jain_fractons_2022,bidussi_fractons_2022}. A resolution of this puzzle is that in the dual language, metric fluctuations are mapped to additional tensor gauge variables.

\section*{Acknowledgments}
J.J.F.-M. and J.M.-V. thank the financial support of Spanish Ministerio de Ciencia e Innovación PID2021-125700NAC22. J.M.-V. also thanks  Programa Recualificaci\'on del Sistema Universitario Español 2021–2023. P.S. acknowledges support form the Polish National Science Centre (NCN) Sonata Bis grant 2019/34/E/ST3/00405. P.S. thanks University of Murcia for hospitality and support from the Talento program. L.T. has been supported in part by the Deutsche Forschungsgemeinschaft through the cluster of excellence ct.qmat (Exzellenzcluster 2147, Project 390858490).

\bibliography{CurvedCrystallography}

\begin{thebibliography}{42}%
\makeatletter
\providecommand \@ifxundefined [1]{%
 \@ifx{#1\undefined}
}%
\providecommand \@ifnum [1]{%
 \ifnum #1\expandafter \@firstoftwo
 \else \expandafter \@secondoftwo
 \fi
}%
\providecommand \@ifx [1]{%
 \ifx #1\expandafter \@firstoftwo
 \else \expandafter \@secondoftwo
 \fi
}%
\providecommand \natexlab [1]{#1}%
\providecommand \enquote  [1]{``#1''}%
\providecommand \bibnamefont  [1]{#1}%
\providecommand \bibfnamefont [1]{#1}%
\providecommand \citenamefont [1]{#1}%
\providecommand \href@noop [0]{\@secondoftwo}%
\providecommand \href [0]{\begingroup \@sanitize@url \@href}%
\providecommand \@href[1]{\@@startlink{#1}\@@href}%
\providecommand \@@href[1]{\endgroup#1\@@endlink}%
\providecommand \@sanitize@url [0]{\catcode `\\12\catcode `\$12\catcode
  `\&12\catcode `\#12\catcode `\^12\catcode `\_12\catcode `\%12\relax}%
\providecommand \@@startlink[1]{}%
\providecommand \@@endlink[0]{}%
\providecommand \url  [0]{\begingroup\@sanitize@url \@url }%
\providecommand \@url [1]{\endgroup\@href {#1}{\urlprefix }}%
\providecommand \urlprefix  [0]{URL }%
\providecommand \Eprint [0]{\href }%
\providecommand \doibase [0]{https://doi.org/}%
\providecommand \selectlanguage [0]{\@gobble}%
\providecommand \bibinfo  [0]{\@secondoftwo}%
\providecommand \bibfield  [0]{\@secondoftwo}%
\providecommand \translation [1]{[#1]}%
\providecommand \BibitemOpen [0]{}%
\providecommand \bibitemStop [0]{}%
\providecommand \bibitemNoStop [0]{.\EOS\space}%
\providecommand \EOS [0]{\spacefactor3000\relax}%
\providecommand \BibitemShut  [1]{\csname bibitem#1\endcsname}%
\let\auto@bib@innerbib\@empty
\bibitem [{\citenamefont {Koning}\ and\ \citenamefont
  {Vitelli}(2016)}]{fernandez-nieves_crystals_2016}%
  \BibitemOpen
  \bibfield  {author} {\bibinfo {author} {\bibfnamefont {V.}~\bibnamefont
  {Koning}}\ and\ \bibinfo {author} {\bibfnamefont {V.}~\bibnamefont
  {Vitelli}},\ }\bibfield  {title} {\bibinfo {title} {Crystals and {Liquid}
  {Crystals} {Confined} to {Curved} {Geometries}},\ }in\ \href
  {https://doi.org/10.1002/9781119220510.ch19} {\emph {\bibinfo {booktitle}
  {Fluids, {Colloids} and {Soft} {Materials}}}},\ \bibinfo {editor} {edited by\
  \bibinfo {editor} {\bibfnamefont {A.}~\bibnamefont {Fernandez-Nieves}}\ and\
  \bibinfo {editor} {\bibfnamefont {A.~M.}\ \bibnamefont {Puertas}}}\ (\bibinfo
   {publisher} {Wiley},\ \bibinfo {address} {Hoboken, NJ},\ \bibinfo {year}
  {2016})\ pp.\ \bibinfo {pages} {369--386}\BibitemShut {NoStop}%
\bibitem [{\citenamefont {Bertoldi}\ \emph {et~al.}(2017)\citenamefont
  {Bertoldi}, \citenamefont {Vitelli}, \citenamefont {Christensen},\ and\
  \citenamefont {van Hecke}}]{bertoldi_flexible_2017}%
  \BibitemOpen
  \bibfield  {author} {\bibinfo {author} {\bibfnamefont {K.}~\bibnamefont
  {Bertoldi}}, \bibinfo {author} {\bibfnamefont {V.}~\bibnamefont {Vitelli}},
  \bibinfo {author} {\bibfnamefont {J.}~\bibnamefont {Christensen}},\ and\
  \bibinfo {author} {\bibfnamefont {M.}~\bibnamefont {van Hecke}},\ }\bibfield
  {title} {\bibinfo {title} {Flexible mechanical metamaterials},\ }\href
  {https://doi.org/10.1038/natrevmats.2017.66} {\bibfield  {journal} {\bibinfo
  {journal} {Nature Reviews Materials}\ }\textbf {\bibinfo {volume} {2}},\
  \bibinfo {pages} {1} (\bibinfo {year} {2017})}\BibitemShut {NoStop}%
\bibitem [{\citenamefont {Lecuit}\ and\ \citenamefont
  {Lenne}(2007)}]{lecuit_cell_2007}%
  \BibitemOpen
  \bibfield  {author} {\bibinfo {author} {\bibfnamefont {T.}~\bibnamefont
  {Lecuit}}\ and\ \bibinfo {author} {\bibfnamefont {P.-F.}\ \bibnamefont
  {Lenne}},\ }\bibfield  {title} {\bibinfo {title} {Cell surface mechanics and
  the control of cell shape, tissue patterns and morphogenesis},\ }\href
  {https://doi.org/10.1038/nrm2222} {\bibfield  {journal} {\bibinfo  {journal}
  {Nature Reviews Molecular Cell Biology}\ }\textbf {\bibinfo {volume} {8}},\
  \bibinfo {pages} {633} (\bibinfo {year} {2007})}\BibitemShut {NoStop}%
\bibitem [{\citenamefont {Kleinert}(1989)}]{kleinert_gauge_1989}%
  \BibitemOpen
  \bibfield  {author} {\bibinfo {author} {\bibfnamefont {H.}~\bibnamefont
  {Kleinert}},\ }\href@noop {} {\emph {\bibinfo {title} {Gauge fields in
  condensed matter}}}\ (\bibinfo  {publisher} {World Scientific},\ \bibinfo
  {address} {Singapore},\ \bibinfo {year} {1989})\BibitemShut {NoStop}%
\bibitem [{\citenamefont {Kleinert}(1982)}]{kleinert_duality_1982}%
  \BibitemOpen
  \bibfield  {author} {\bibinfo {author} {\bibfnamefont {H.}~\bibnamefont
  {Kleinert}},\ }\bibfield  {title} {\bibinfo {title} {Duality transformation
  for defect melting},\ }\href {https://doi.org/10.1016/0375-9601(82)90578-3}
  {\bibfield  {journal} {\bibinfo  {journal} {Physics Letters A}\ }\textbf
  {\bibinfo {volume} {91}},\ \bibinfo {pages} {295} (\bibinfo {year}
  {1982})}\BibitemShut {NoStop}%
\bibitem [{\citenamefont {Kleinert}(1983)}]{kleinert_double_1983}%
  \BibitemOpen
  \bibfield  {author} {\bibinfo {author} {\bibfnamefont {H.}~\bibnamefont
  {Kleinert}},\ }\bibfield  {title} {\bibinfo {title} {Double gauge theory of
  stresses and defects},\ }\href {https://doi.org/10.1016/0375-9601(83)90099-3}
  {\bibfield  {journal} {\bibinfo  {journal} {Physics Letters A}\ }\textbf
  {\bibinfo {volume} {97}},\ \bibinfo {pages} {51} (\bibinfo {year}
  {1983})}\BibitemShut {NoStop}%
\bibitem [{\citenamefont {Beekman}\ \emph
  {et~al.}(2017{\natexlab{a}})\citenamefont {Beekman}, \citenamefont
  {Nissinen}, \citenamefont {Wu},\ and\ \citenamefont
  {Zaanen}}]{beekman_dual_2017-1}%
  \BibitemOpen
  \bibfield  {author} {\bibinfo {author} {\bibfnamefont {A.~J.}\ \bibnamefont
  {Beekman}}, \bibinfo {author} {\bibfnamefont {J.}~\bibnamefont {Nissinen}},
  \bibinfo {author} {\bibfnamefont {K.}~\bibnamefont {Wu}},\ and\ \bibinfo
  {author} {\bibfnamefont {J.}~\bibnamefont {Zaanen}},\ }\bibfield  {title}
  {\bibinfo {title} {Dual gauge field theory of quantum liquid crystals in
  three dimensions},\ }\href {https://doi.org/10.1103/PhysRevB.96.165115}
  {\bibfield  {journal} {\bibinfo  {journal} {Physical Review B}\ }\textbf
  {\bibinfo {volume} {96}},\ \bibinfo {pages} {165115} (\bibinfo {year}
  {2017}{\natexlab{a}})}\BibitemShut {NoStop}%
\bibitem [{\citenamefont {Beekman}\ \emph
  {et~al.}(2017{\natexlab{b}})\citenamefont {Beekman}, \citenamefont
  {Nissinen}, \citenamefont {Wu}, \citenamefont {Liu}, \citenamefont {Slager},
  \citenamefont {Nussinov}, \citenamefont {Cvetkovic},\ and\ \citenamefont
  {Zaanen}}]{beekman_dual_2017}%
  \BibitemOpen
  \bibfield  {author} {\bibinfo {author} {\bibfnamefont {A.}~\bibnamefont
  {Beekman}}, \bibinfo {author} {\bibfnamefont {J.}~\bibnamefont {Nissinen}},
  \bibinfo {author} {\bibfnamefont {K.}~\bibnamefont {Wu}}, \bibinfo {author}
  {\bibfnamefont {K.}~\bibnamefont {Liu}}, \bibinfo {author} {\bibfnamefont
  {R.-J.}\ \bibnamefont {Slager}}, \bibinfo {author} {\bibfnamefont
  {Z.}~\bibnamefont {Nussinov}}, \bibinfo {author} {\bibfnamefont
  {V.}~\bibnamefont {Cvetkovic}},\ and\ \bibinfo {author} {\bibfnamefont
  {J.}~\bibnamefont {Zaanen}},\ }\bibfield  {title} {\bibinfo {title} {Dual
  gauge field theory of quantum liquid crystals in two dimensions},\ }\href
  {https://doi.org/10.1016/j.physrep.2017.03.004} {\bibfield  {journal}
  {\bibinfo  {journal} {Physics Reports}\ }\textbf {\bibinfo {volume} {683}},\
  \bibinfo {pages} {1} (\bibinfo {year} {2017}{\natexlab{b}})}\BibitemShut
  {NoStop}%
\bibitem [{\citenamefont {Pretko}\ and\ \citenamefont
  {Radzihovsky}(2018)}]{pretko_fracton-elasticity_2018}%
  \BibitemOpen
  \bibfield  {author} {\bibinfo {author} {\bibfnamefont {M.}~\bibnamefont
  {Pretko}}\ and\ \bibinfo {author} {\bibfnamefont {L.}~\bibnamefont
  {Radzihovsky}},\ }\bibfield  {title} {\bibinfo {title} {Fracton-{Elasticity}
  {Duality}},\ }\href {https://doi.org/10.1103/PhysRevLett.120.195301}
  {\bibfield  {journal} {\bibinfo  {journal} {Physical Review Letters}\
  }\textbf {\bibinfo {volume} {120}},\ \bibinfo {pages} {195301} (\bibinfo
  {year} {2018})}\BibitemShut {NoStop}%
\bibitem [{\citenamefont {Pretko}\ \emph {et~al.}(2019)\citenamefont {Pretko},
  \citenamefont {Zhai},\ and\ \citenamefont
  {Radzihovsky}}]{pretko_crystal--fracton_2019}%
  \BibitemOpen
  \bibfield  {author} {\bibinfo {author} {\bibfnamefont {M.}~\bibnamefont
  {Pretko}}, \bibinfo {author} {\bibfnamefont {Z.}~\bibnamefont {Zhai}},\ and\
  \bibinfo {author} {\bibfnamefont {L.}~\bibnamefont {Radzihovsky}},\
  }\bibfield  {title} {\bibinfo {title} {Crystal-to-fracton tensor gauge theory
  dualities},\ }\href {https://doi.org/10.1103/PhysRevB.100.134113} {\bibfield
  {journal} {\bibinfo  {journal} {Physical Review B}\ }\textbf {\bibinfo
  {volume} {100}},\ \bibinfo {pages} {134113} (\bibinfo {year}
  {2019})}\BibitemShut {NoStop}%
\bibitem [{\citenamefont {Radzihovsky}\ and\ \citenamefont
  {Hermele}(2020)}]{radzihovsky_fractons_2020}%
  \BibitemOpen
  \bibfield  {author} {\bibinfo {author} {\bibfnamefont {L.}~\bibnamefont
  {Radzihovsky}}\ and\ \bibinfo {author} {\bibfnamefont {M.}~\bibnamefont
  {Hermele}},\ }\bibfield  {title} {\bibinfo {title} {Fractons from {Vector}
  {Gauge} {Theory}},\ }\href {https://doi.org/10.1103/PhysRevLett.124.050402}
  {\bibfield  {journal} {\bibinfo  {journal} {Physical Review Letters}\
  }\textbf {\bibinfo {volume} {124}},\ \bibinfo {pages} {050402} (\bibinfo
  {year} {2020})}\BibitemShut {NoStop}%
\bibitem [{\citenamefont {Radzihovsky}(2020)}]{radzihovsky_quantum_2020}%
  \BibitemOpen
  \bibfield  {author} {\bibinfo {author} {\bibfnamefont {L.}~\bibnamefont
  {Radzihovsky}},\ }\bibfield  {title} {\bibinfo {title} {Quantum {Smectic}
  {Gauge} {Theory}},\ }\href {https://doi.org/10.1103/PhysRevLett.125.267601}
  {\bibfield  {journal} {\bibinfo  {journal} {Physical Review Letters}\
  }\textbf {\bibinfo {volume} {125}},\ \bibinfo {pages} {267601} (\bibinfo
  {year} {2020})}\BibitemShut {NoStop}%
\bibitem [{\citenamefont {Gromov}\ and\ \citenamefont
  {Sur\'{o}wka}(2020)}]{gromov_duality_2020}%
  \BibitemOpen
  \bibfield  {author} {\bibinfo {author} {\bibfnamefont {A.}~\bibnamefont
  {Gromov}}\ and\ \bibinfo {author} {\bibfnamefont {P.}~\bibnamefont
  {Sur\'{o}wka}},\ }\bibfield  {title} {\bibinfo {title} {On duality between
  {Cosserat} elasticity and fractons},\ }\href
  {https://doi.org/10.21468/SciPostPhys.8.4.065} {\bibfield  {journal}
  {\bibinfo  {journal} {SciPost Physics}\ }\textbf {\bibinfo {volume} {8}},\
  \bibinfo {pages} {065} (\bibinfo {year} {2020})}\BibitemShut {NoStop}%
\bibitem [{\citenamefont {Sur\'{o}wka}(2021)}]{surowka_dual_2021}%
  \BibitemOpen
  \bibfield  {author} {\bibinfo {author} {\bibfnamefont {P.}~\bibnamefont
  {Sur\'{o}wka}},\ }\bibfield  {title} {\bibinfo {title} {Dual gauge theory
  formulation of planar quasicrystal elasticity and fractons},\ }\href
  {https://doi.org/10.1103/PhysRevB.103.L201119} {\bibfield  {journal}
  {\bibinfo  {journal} {Physical Review B}\ }\textbf {\bibinfo {volume}
  {103}},\ \bibinfo {pages} {L201119} (\bibinfo {year} {2021})}\BibitemShut
  {NoStop}%
\bibitem [{\citenamefont {Hirono}\ and\ \citenamefont
  {Qi}(2022)}]{hirono_effective_2022}%
  \BibitemOpen
  \bibfield  {author} {\bibinfo {author} {\bibfnamefont {Y.}~\bibnamefont
  {Hirono}}\ and\ \bibinfo {author} {\bibfnamefont {Y.-H.}\ \bibnamefont
  {Qi}},\ }\bibfield  {title} {\bibinfo {title} {Effective field theories for
  gapless phases with fractons via a coset construction},\ }\href
  {https://doi.org/10.1103/PhysRevB.105.205109} {\bibfield  {journal} {\bibinfo
   {journal} {Physical Review B}\ }\textbf {\bibinfo {volume} {105}},\ \bibinfo
  {pages} {205109} (\bibinfo {year} {2022})}\BibitemShut {NoStop}%
\bibitem [{\citenamefont {Caddeo}\ \emph {et~al.}(2022)\citenamefont {Caddeo},
  \citenamefont {Hoyos},\ and\ \citenamefont {Musso}}]{caddeo_emergent_2022}%
  \BibitemOpen
  \bibfield  {author} {\bibinfo {author} {\bibfnamefont {A.}~\bibnamefont
  {Caddeo}}, \bibinfo {author} {\bibfnamefont {C.}~\bibnamefont {Hoyos}},\ and\
  \bibinfo {author} {\bibfnamefont {D.}~\bibnamefont {Musso}},\ }\bibfield
  {title} {\bibinfo {title} {Emergent dipole gauge fields and fractons},\
  }\href {https://doi.org/10.1103/PhysRevD.106.L111903} {\bibfield  {journal}
  {\bibinfo  {journal} {Physical Review D}\ }\textbf {\bibinfo {volume}
  {106}},\ \bibinfo {pages} {L111903} (\bibinfo {year} {2022})}\BibitemShut
  {NoStop}%
\bibitem [{\citenamefont {Beekman}\ \emph {et~al.}(2022)\citenamefont
  {Beekman}, \citenamefont {Balm},\ and\ \citenamefont
  {Zaanen}}]{zaanen_crystal_2022}%
  \BibitemOpen
  \bibfield  {author} {\bibinfo {author} {\bibfnamefont {A.}~\bibnamefont
  {Beekman}}, \bibinfo {author} {\bibfnamefont {F.}~\bibnamefont {Balm}},\ and\
  \bibinfo {author} {\bibfnamefont {J.}~\bibnamefont {Zaanen}},\ }\bibfield
  {title} {\bibinfo {title} {Crystal gravity},\ }\href
  {https://doi.org/10.21468/SciPostPhys.13.2.039} {\bibfield  {journal}
  {\bibinfo  {journal} {SciPost Physics}\ }\textbf {\bibinfo {volume} {13}},\
  \bibinfo {pages} {039} (\bibinfo {year} {2022})}\BibitemShut {NoStop}%
\bibitem [{\citenamefont {Kleinert}(2008)}]{kleinert_multivalued_2008}%
  \BibitemOpen
  \bibfield  {author} {\bibinfo {author} {\bibfnamefont {H.}~\bibnamefont
  {Kleinert}},\ }\href@noop {} {\emph {\bibinfo {title} {Multivalued
  {Fields}}}}\ (\bibinfo  {publisher} {World Scientific},\ \bibinfo {address}
  {Singapore},\ \bibinfo {year} {2008})\BibitemShut {NoStop}%
\bibitem [{\citenamefont {Efrati}\ \emph {et~al.}(2009)\citenamefont {Efrati},
  \citenamefont {Sharon},\ and\ \citenamefont
  {Kupferman}}]{efrati_elastic_2009}%
  \BibitemOpen
  \bibfield  {author} {\bibinfo {author} {\bibfnamefont {E.}~\bibnamefont
  {Efrati}}, \bibinfo {author} {\bibfnamefont {E.}~\bibnamefont {Sharon}},\
  and\ \bibinfo {author} {\bibfnamefont {R.}~\bibnamefont {Kupferman}},\
  }\bibfield  {title} {\bibinfo {title} {Elastic theory of unconstrained
  non-{Euclidean} plates},\ }\href {https://doi.org/10.1016/j.jmps.2008.12.004}
  {\bibfield  {journal} {\bibinfo  {journal} {Journal of the Mechanics and
  Physics of Solids}\ }\textbf {\bibinfo {volume} {57}},\ \bibinfo {pages}
  {762} (\bibinfo {year} {2009})}\BibitemShut {NoStop}%
\bibitem [{\citenamefont {Moshe}\ \emph {et~al.}(2015)\citenamefont {Moshe},
  \citenamefont {Sharon},\ and\ \citenamefont
  {Kupferman}}]{moshe_elastic_2015}%
  \BibitemOpen
  \bibfield  {author} {\bibinfo {author} {\bibfnamefont {M.}~\bibnamefont
  {Moshe}}, \bibinfo {author} {\bibfnamefont {E.}~\bibnamefont {Sharon}},\ and\
  \bibinfo {author} {\bibfnamefont {R.}~\bibnamefont {Kupferman}},\ }\bibfield
  {title} {\bibinfo {title} {Elastic interactions between two-dimensional
  geometric defects},\ }\href {https://doi.org/10.1103/PhysRevE.92.062403}
  {\bibfield  {journal} {\bibinfo  {journal} {Physical Review E}\ }\textbf
  {\bibinfo {volume} {92}},\ \bibinfo {pages} {062403} (\bibinfo {year}
  {2015})}\BibitemShut {NoStop}%
\bibitem [{\citenamefont {Li}\ \emph {et~al.}(2019)\citenamefont {Li},
  \citenamefont {Zandi},\ and\ \citenamefont {Travesset}}]{li_elasticity_2019}%
  \BibitemOpen
  \bibfield  {author} {\bibinfo {author} {\bibfnamefont {S.}~\bibnamefont
  {Li}}, \bibinfo {author} {\bibfnamefont {R.}~\bibnamefont {Zandi}},\ and\
  \bibinfo {author} {\bibfnamefont {A.}~\bibnamefont {Travesset}},\ }\bibfield
  {title} {\bibinfo {title} {Elasticity in curved topographies: {Exact}
  theories and linear approximations},\ }\href
  {https://doi.org/10.1103/PhysRevE.99.063005} {\bibfield  {journal} {\bibinfo
  {journal} {Physical Review E}\ }\textbf {\bibinfo {volume} {99}},\ \bibinfo
  {pages} {063005} (\bibinfo {year} {2019})}\BibitemShut {NoStop}%
\bibitem [{\citenamefont {Ciarlet}(2000)}]{ciarlet}%
  \BibitemOpen
  \bibfield  {author} {\bibinfo {author} {\bibfnamefont {P.~G.}\ \bibnamefont
  {Ciarlet}},\ }\href@noop {} {\emph {\bibinfo {title} {Theory of Shells}}}\
  (\bibinfo  {publisher} {Elsevier},\ \bibinfo {address} {Amsterdam},\ \bibinfo
  {year} {2000})\BibitemShut {NoStop}%
\bibitem [{\citenamefont {Carter}\ and\ \citenamefont
  {Quintana}(1972)}]{Carter_Quintana_foundations_1972}%
  \BibitemOpen
  \bibfield  {author} {\bibinfo {author} {\bibfnamefont {B.}~\bibnamefont
  {Carter}}\ and\ \bibinfo {author} {\bibfnamefont {H.}~\bibnamefont
  {Quintana}},\ }\bibfield  {title} {\bibinfo {title} {Foundations of general
  relativistic high-pressure elasticity theory},\ }\href
  {https://doi.org/10.1098/rspa.1972.0164} {\bibfield  {journal} {\bibinfo
  {journal} {Proceedings of the Royal Society of London. A. Mathematical and
  Physical Sciences}\ }\textbf {\bibinfo {volume} {331}},\ \bibinfo {pages}
  {57} (\bibinfo {year} {1972})}\BibitemShut {NoStop}%
\bibitem [{\citenamefont {Beig}\ and\ \citenamefont
  {Schmidt}(2003)}]{beig_relativistic_2003}%
  \BibitemOpen
  \bibfield  {author} {\bibinfo {author} {\bibfnamefont {R.}~\bibnamefont
  {Beig}}\ and\ \bibinfo {author} {\bibfnamefont {B.~G.}\ \bibnamefont
  {Schmidt}},\ }\bibfield  {title} {\bibinfo {title} {Relativistic
  elasticity},\ }\href {https://doi.org/10.1088/0264-9381/20/5/308} {\bibfield
  {journal} {\bibinfo  {journal} {Classical and Quantum Gravity}\ }\textbf
  {\bibinfo {volume} {20}},\ \bibinfo {pages} {889} (\bibinfo {year}
  {2003})}\BibitemShut {NoStop}%
\bibitem [{\citenamefont {Kamien}(2002)}]{kamien_geometry_2002}%
  \BibitemOpen
  \bibfield  {author} {\bibinfo {author} {\bibfnamefont {R.~D.}\ \bibnamefont
  {Kamien}},\ }\bibfield  {title} {\bibinfo {title} {The geometry of soft
  materials: a primer},\ }\href {https://doi.org/10.1103/RevModPhys.74.953}
  {\bibfield  {journal} {\bibinfo  {journal} {Reviews of Modern Physics}\
  }\textbf {\bibinfo {volume} {74}},\ \bibinfo {pages} {953} (\bibinfo {year}
  {2002})}\BibitemShut {NoStop}%
\bibitem [{\citenamefont {Vitelli}\ \emph {et~al.}(2006)\citenamefont
  {Vitelli}, \citenamefont {Lucks},\ and\ \citenamefont
  {Nelson}}]{vitelli_crystallography_2006}%
  \BibitemOpen
  \bibfield  {author} {\bibinfo {author} {\bibfnamefont {V.}~\bibnamefont
  {Vitelli}}, \bibinfo {author} {\bibfnamefont {J.~B.}\ \bibnamefont {Lucks}},\
  and\ \bibinfo {author} {\bibfnamefont {D.~R.}\ \bibnamefont {Nelson}},\
  }\bibfield  {title} {\bibinfo {title} {Crystallography on curved surfaces},\
  }\href {https://doi.org/10.1073/pnas.0602755103} {\bibfield  {journal}
  {\bibinfo  {journal} {Proceedings of the National Academy of Sciences}\
  }\textbf {\bibinfo {volume} {103}},\ \bibinfo {pages} {12323} (\bibinfo
  {year} {2006})}\BibitemShut {NoStop}%
\bibitem [{\citenamefont {Bowick}\ and\ \citenamefont
  {Giomi}(2009)}]{bowick_two-dimensional_2009}%
  \BibitemOpen
  \bibfield  {author} {\bibinfo {author} {\bibfnamefont {M.~J.}\ \bibnamefont
  {Bowick}}\ and\ \bibinfo {author} {\bibfnamefont {L.}~\bibnamefont {Giomi}},\
  }\bibfield  {title} {\bibinfo {title} {Two-dimensional matter: order,
  curvature and defects},\ }\href {https://doi.org/10.1080/00018730903043166}
  {\bibfield  {journal} {\bibinfo  {journal} {Advances in Physics}\ }\textbf
  {\bibinfo {volume} {58}},\ \bibinfo {pages} {449} (\bibinfo {year}
  {2009})}\BibitemShut {NoStop}%
\bibitem [{\citenamefont {Paczuski}\ \emph {et~al.}(1988)\citenamefont
  {Paczuski}, \citenamefont {Kardar},\ and\ \citenamefont
  {Nelson}}]{paczuski_landau_1988}%
  \BibitemOpen
  \bibfield  {author} {\bibinfo {author} {\bibfnamefont {M.}~\bibnamefont
  {Paczuski}}, \bibinfo {author} {\bibfnamefont {M.}~\bibnamefont {Kardar}},\
  and\ \bibinfo {author} {\bibfnamefont {D.~R.}\ \bibnamefont {Nelson}},\
  }\bibfield  {title} {\bibinfo {title} {Landau {Theory} of the {Crumpling}
  {Transition}},\ }\href {https://doi.org/10.1103/PhysRevLett.60.2638}
  {\bibfield  {journal} {\bibinfo  {journal} {Physical Review Letters}\
  }\textbf {\bibinfo {volume} {60}},\ \bibinfo {pages} {2638} (\bibinfo {year}
  {1988})}\BibitemShut {NoStop}%
\bibitem [{\citenamefont {Nelson}(2002)}]{nelson_defects_2002}%
  \BibitemOpen
  \bibfield  {author} {\bibinfo {author} {\bibfnamefont {D.~R.}\ \bibnamefont
  {Nelson}},\ }\href@noop {} {\emph {\bibinfo {title} {Defects and geometry in
  condensed matter physics}}}\ (\bibinfo  {publisher} {Cambridge University
  Press},\ \bibinfo {address} {Cambridge},\ \bibinfo {year} {2002})\BibitemShut
  {NoStop}%
\bibitem [{\citenamefont {Fierz}\ and\ \citenamefont
  {Pauli}(1939)}]{Fierz_Pauli_1939}%
  \BibitemOpen
  \bibfield  {author} {\bibinfo {author} {\bibfnamefont {M.}~\bibnamefont
  {Fierz}}\ and\ \bibinfo {author} {\bibfnamefont {W.~E.}\ \bibnamefont
  {Pauli}},\ }\bibfield  {title} {\bibinfo {title} {On relativistic wave
  equations for particles of arbitrary spin in an electromagnetic field},\
  }\href {https://doi.org/10.1098/rspa.1939.0140} {\bibfield  {journal}
  {\bibinfo  {journal} {Proceedings of the Royal Society of London. Series A.
  Mathematical and Physical Sciences}\ }\textbf {\bibinfo {volume} {173}},\
  \bibinfo {pages} {211} (\bibinfo {year} {1939})}\BibitemShut {NoStop}%
\bibitem [{\citenamefont {Hamber}(2008)}]{hamber_quantum_2008}%
  \BibitemOpen
  \bibfield  {author} {\bibinfo {author} {\bibfnamefont {H.~W.}\ \bibnamefont
  {Hamber}},\ }\href {https://doi.org/10.1007/978-3-540-85293-3} {\emph
  {\bibinfo {title} {Quantum {Gravitation}}}}\ (\bibinfo  {publisher}
  {Springer},\ \bibinfo {address} {Berlin},\ \bibinfo {year}
  {2008})\BibitemShut {NoStop}%
\bibitem [{\citenamefont {Witten}(1988)}]{witten_2_1988}%
  \BibitemOpen
  \bibfield  {author} {\bibinfo {author} {\bibfnamefont {E.}~\bibnamefont
  {Witten}},\ }\bibfield  {title} {\bibinfo {title} {2 + 1 dimensional gravity
  as an exactly soluble system},\ }\href
  {https://doi.org/10.1016/0550-3213(88)90143-5} {\bibfield  {journal}
  {\bibinfo  {journal} {Nuclear Physics B}\ }\textbf {\bibinfo {volume}
  {311}},\ \bibinfo {pages} {46} (\bibinfo {year} {1988})}\BibitemShut
  {NoStop}%
\bibitem [{\citenamefont {Utiyama}(1956)}]{utiyama_invariant_1956}%
  \BibitemOpen
  \bibfield  {author} {\bibinfo {author} {\bibfnamefont {R.}~\bibnamefont
  {Utiyama}},\ }\bibfield  {title} {\bibinfo {title} {Invariant {Theoretical}
  {Interpretation} of {Interaction}},\ }\href
  {https://doi.org/10.1103/PhysRev.101.1597} {\bibfield  {journal} {\bibinfo
  {journal} {Physical Review}\ }\textbf {\bibinfo {volume} {101}},\ \bibinfo
  {pages} {1597} (\bibinfo {year} {1956})}\BibitemShut {NoStop}%
\bibitem [{\citenamefont {Kibble}(1961)}]{kibble_lorentz_1961}%
  \BibitemOpen
  \bibfield  {author} {\bibinfo {author} {\bibfnamefont {T.~W.~B.}\
  \bibnamefont {Kibble}},\ }\bibfield  {title} {\bibinfo {title} {Lorentz
  {Invariance} and the {Gravitational} {Field}},\ }\href
  {https://doi.org/10.1063/1.1703702} {\bibfield  {journal} {\bibinfo
  {journal} {Journal of Mathematical Physics}\ }\textbf {\bibinfo {volume}
  {2}},\ \bibinfo {pages} {212} (\bibinfo {year} {1961})}\BibitemShut {NoStop}%
\bibitem [{\citenamefont {Sciama}(1962)}]{sciama}%
  \BibitemOpen
  \bibfield  {author} {\bibinfo {author} {\bibfnamefont {D.~W.}\ \bibnamefont
  {Sciama}},\ }\bibfield  {title} {\bibinfo {title} {On the analogy between
  charge and spin in general relativity},\ }in\ \href@noop {} {\emph {\bibinfo
  {booktitle} {Recent Developments in General Relativity}}}\ (\bibinfo
  {publisher} {Polish Scientific Publishers},\ \bibinfo {address} {Warsaw},\
  \bibinfo {year} {1962})\BibitemShut {NoStop}%
\bibitem [{\citenamefont {Ivanov}\ and\ \citenamefont
  {Niederle}(1982)}]{ivanov_gauge_1982}%
  \BibitemOpen
  \bibfield  {author} {\bibinfo {author} {\bibfnamefont {E.~A.}\ \bibnamefont
  {Ivanov}}\ and\ \bibinfo {author} {\bibfnamefont {J.}~\bibnamefont
  {Niederle}},\ }\bibfield  {title} {\bibinfo {title} {Gauge formulation of
  gravitation theories. {I}. {The} {Poincar\'{e}}, de {Sitter}, and conformal
  cases},\ }\href {https://doi.org/10.1103/PhysRevD.25.976} {\bibfield
  {journal} {\bibinfo  {journal} {Physical Review D}\ }\textbf {\bibinfo
  {volume} {25}},\ \bibinfo {pages} {976} (\bibinfo {year} {1982})}\BibitemShut
  {NoStop}%
\bibitem [{\citenamefont {Grignani}\ and\ \citenamefont
  {Nardelli}(1992)}]{grignani_gravity_1992}%
  \BibitemOpen
  \bibfield  {author} {\bibinfo {author} {\bibfnamefont {G.}~\bibnamefont
  {Grignani}}\ and\ \bibinfo {author} {\bibfnamefont {G.}~\bibnamefont
  {Nardelli}},\ }\bibfield  {title} {\bibinfo {title} {Gravity and the
  {Poincar\'{e}} group},\ }\href {https://doi.org/10.1103/PhysRevD.45.2719}
  {\bibfield  {journal} {\bibinfo  {journal} {Physical Review D}\ }\textbf
  {\bibinfo {volume} {45}},\ \bibinfo {pages} {2719} (\bibinfo {year}
  {1992})}\BibitemShut {NoStop}%
\bibitem [{\citenamefont {Peña-Benítez}(2023)}]{pena-benitez_fractons_2023}%
  \BibitemOpen
  \bibfield  {author} {\bibinfo {author} {\bibfnamefont {F.}~\bibnamefont
  {Peña-Benítez}},\ }\bibfield  {title} {\bibinfo {title} {Fractons,
  symmetric gauge fields and geometry},\ }\href
  {https://doi.org/10.1103/PhysRevResearch.5.013101} {\bibfield  {journal}
  {\bibinfo  {journal} {Physical Review Research}\ }\textbf {\bibinfo {volume}
  {5}},\ \bibinfo {pages} {013101} (\bibinfo {year} {2023})}\BibitemShut
  {NoStop}%
\bibitem [{\citenamefont {Cvetkovic}\ \emph {et~al.}(2006)\citenamefont
  {Cvetkovic}, \citenamefont {Nussinov},\ and\ \citenamefont
  {Zaanen}}]{cvetkovic_topological_2006}%
  \BibitemOpen
  \bibfield  {author} {\bibinfo {author} {\bibfnamefont {V.}~\bibnamefont
  {Cvetkovic}}, \bibinfo {author} {\bibfnamefont {Z.}~\bibnamefont
  {Nussinov}},\ and\ \bibinfo {author} {\bibfnamefont {J.}~\bibnamefont
  {Zaanen}},\ }\bibfield  {title} {\bibinfo {title} {Topological kinematic
  constraints: dislocations and the glide principle},\ }\href
  {https://doi.org/10.1080/14786430600636328} {\bibfield  {journal} {\bibinfo
  {journal} {Philosophical Magazine}\ }\textbf {\bibinfo {volume} {86}},\
  \bibinfo {pages} {2995} (\bibinfo {year} {2006})}\BibitemShut {NoStop}%
\bibitem [{\citenamefont {Slagle}\ \emph {et~al.}(2019)\citenamefont {Slagle},
  \citenamefont {Prem},\ and\ \citenamefont {Pretko}}]{slagle_symmetric_2019}%
  \BibitemOpen
  \bibfield  {author} {\bibinfo {author} {\bibfnamefont {K.}~\bibnamefont
  {Slagle}}, \bibinfo {author} {\bibfnamefont {A.}~\bibnamefont {Prem}},\ and\
  \bibinfo {author} {\bibfnamefont {M.}~\bibnamefont {Pretko}},\ }\bibfield
  {title} {\bibinfo {title} {Symmetric tensor gauge theories on curved
  spaces},\ }\href {https://doi.org/10.1016/j.aop.2019.167910} {\bibfield
  {journal} {\bibinfo  {journal} {Annals of Physics}\ }\textbf {\bibinfo
  {volume} {410}},\ \bibinfo {pages} {167910} (\bibinfo {year}
  {2019})}\BibitemShut {NoStop}%
\bibitem [{\citenamefont {Jain}\ and\ \citenamefont
  {Jensen}(2022)}]{jain_fractons_2022}%
  \BibitemOpen
  \bibfield  {author} {\bibinfo {author} {\bibfnamefont {A.}~\bibnamefont
  {Jain}}\ and\ \bibinfo {author} {\bibfnamefont {K.}~\bibnamefont {Jensen}},\
  }\bibfield  {title} {\bibinfo {title} {Fractons in curved space},\ }\href
  {https://doi.org/10.21468/SciPostPhys.12.4.142} {\bibfield  {journal}
  {\bibinfo  {journal} {SciPost Physics}\ }\textbf {\bibinfo {volume} {12}},\
  \bibinfo {pages} {142} (\bibinfo {year} {2022})}\BibitemShut {NoStop}%
\bibitem [{\citenamefont {Bidussi}\ \emph {et~al.}(2022)\citenamefont
  {Bidussi}, \citenamefont {Hartong}, \citenamefont {Have}, \citenamefont
  {Musaeus},\ and\ \citenamefont {Prohazka}}]{bidussi_fractons_2022}%
  \BibitemOpen
  \bibfield  {author} {\bibinfo {author} {\bibfnamefont {L.}~\bibnamefont
  {Bidussi}}, \bibinfo {author} {\bibfnamefont {J.}~\bibnamefont {Hartong}},
  \bibinfo {author} {\bibfnamefont {E.}~\bibnamefont {Have}}, \bibinfo {author}
  {\bibfnamefont {J.}~\bibnamefont {Musaeus}},\ and\ \bibinfo {author}
  {\bibfnamefont {S.}~\bibnamefont {Prohazka}},\ }\bibfield  {title} {\bibinfo
  {title} {Fractons, dipole symmetries and curved spacetime},\ }\href
  {https://doi.org/10.21468/SciPostPhys.12.6.205} {\bibfield  {journal}
  {\bibinfo  {journal} {SciPost Physics}\ }\textbf {\bibinfo {volume} {12}},\
  \bibinfo {pages} {205} (\bibinfo {year} {2022})}\BibitemShut {NoStop}%
\end{thebibliography}%

\onecolumngrid
\appendix
\section{Hubbard Stratonovich Transformation}
The theory under consideration is given by
\begin{align}
    \mathcal{Z}=\int D[\Phi] \, \exp\left(-S\right),\, 
\end{align}
with $D[\Phi]\equiv D[w_{ij},\partial_t u_i, \partial_t h_{ij}, \partial_k h_{ij}]$, the measure of the fields entering the action of the theory
\begin{equation}
    S=\frac{1}{2}\int d^2xdt\left[\rho_d(\partial_tu_i)^2+(\partial_t h_{ij})^2-(u_{ij}+h_{ij})C_{ijkl}(u_{kl}+h_{kl})-\kappa(\partial_kh_{ij})^2\right],
\end{equation}
where $w_{ij}=u_{ij}+h_{ij}$ is the generalized symmetric strain tensor and $u_{ij}=\frac{1}{2}(\partial_iu_j+\partial_ju_i)$ the usual symmetric elasticity strain tensor. The constants $\rho_d$ represent the material 2d density and $b$ acts as an elastic modulus for the gravity part respectively.

Through a Hubbard-Stratonovich transformation we get:
\begin{equation*}
\begin{split}
\mathcal{Z}_{\text{HS}}=\frac{1}{N}\int\, D[\widetilde{\Phi}\, , \Phi]\, \exp(-S_{\text{HS}}),
\end{split}    
\end{equation*}
with $ D[\widetilde{\Phi}\, , \Phi]\equiv D[\pi_i, L_{ij}, T_{ij}, K_{kij}, w_{ij}, \partial_t u_i, \partial_t h_{ij}, \partial_k h_{ij}]$ being the measure in terms of the fields and the HS-conjugate fields and
\begin{equation}
S_{\text{HS}}=\frac{1}{2}\int d^2xdt\left[T_{ij}C^{-1}_{ijkl}T_{kl}-\frac{1}{\rho_d}(\pi_i)^2+\frac{1}{\kappa}(K_{kij})^2-(L_{ij})^2+2\pi_i\partial_t u_i+2L_{ij}\partial_th_{ij}-2w_{ij}T_{ij}-2\partial_kh_{ij}K_{kij}\right] ,\end{equation}
where we took $T_{0j}=\pi_j$ and $K_{0ij}=L_{ij}$. 

\section{Dual Gauge Theory}
To introduce defects in the theory, one splits the fields $u_i=\overline{u}_i+u^{(s)}_i$ and $h_{ij}=\overline{h}_{ij}+h^{(s)}_{ij}$, into a smooth part represented by the bar, and $(s)$ a singular part. Plugging this into the action and carrying out integration by parts on the smooth fields $\bar{u}_i$ and $\bar{h}_{ij}$ one gets
\begin{align}
    S^*_{\text{HS}}&=\frac{1}{2}\int d^2xdt\left[T_{ij}C^{-1}_{ijkl}T_{kl}-\frac{1}{\rho_d}(\pi_i)^2+\frac{1}{\kappa}(K_{kij})^2-(L_{ij})^2\right] + S_{\rm source},\\ 
    S_{\rm source} &= \int d^2xdt\left[\pi_i\partial_t u^{(s)}_i+L_{ij}\partial_th^{(s)}_{ij}-(u^{(s)}_{ij}+h^{(s)}_{ij})T_{ij}-\partial_kh^{(s)}_{ij}K_{kij}\right]
\end{align}
  from which two conservation laws can be derived. The first one is the usual Hooke's law  
\begin{equation}
\partial_t\pi_i-\partial_jT_{ij}=0 \, .
\label{eq:hookes}
\end{equation}
This equation can be resolved in terms of set of gauge potentials $(\phi, A_{ij})$ as:
\begin{align}
\pi_i&=-\epsilon_{ij}\epsilon_{kl}\partial_kA_{lj}=-\epsilon_{ij}B_j\, , \\ \nonumber
T_{ij}&=-\epsilon_{ik}\epsilon_{jl}(\partial_tA_{kl}+\partial_k\partial_l\phi)=\epsilon_{ik}\epsilon_{jl}E_{kl}  \, , 
\end{align}
where 
\begin{equation}
B_i=\epsilon_{jk}\partial_jA_{ki}\, ,\qquad 
E_{ij}=-\partial_tA_{ij}-\partial_i\partial_j \phi\, ,
\end{equation}
with $E_{ij}$ being the dual electric tensor field and $B_i$ the dual magnetic vector field.

The second conservation law is given by
\begin{equation}
\partial_tL_{ij}-\partial_kK_{kij}+T_{ij}=0\, , 
\end{equation}
which, written in terms of the gauge fields introduced above, can be written as
\begin{equation}
    \partial_t(L_{ij}-\epsilon_{ik}\epsilon_{jl}A_{kl})-\partial_k(K_{kij}+\epsilon_{ik}\epsilon_{jl}\partial_l\phi)=0\, .
\end{equation}
This second  equation can be resolved by a second set of gauge fields $(\Tilde{\phi}_{kl}, \Tilde{A}_{klm})$
\begin{align}
\Tilde{B}_{ij}&=L_{ij}-\epsilon_{ik}\epsilon_{jl}A_{kl}=\epsilon_{ik}\epsilon_{jl}\partial_m\Tilde{A}_{mkl}\, ,\\ \nonumber
\Tilde{E}_{mij}&=K_{mij}+\epsilon_{im}\epsilon_{jl}\partial_l\phi=\epsilon_{ik}\epsilon_{jl}(\partial_t\Tilde{A}_{mkl}+\epsilon_{mn}\partial_n\Tilde{\phi}_{kl})\, ,
\end{align}
which allows one to write it as a Faraday type equation as 
\begin{equation}
     \partial_t\, \Tilde{B}_{ij}-\partial_k\, \Tilde{E}_{kij}=0,
\end{equation}
where now $\Tilde{E}_{mij}$ is the dual rank-3 tensor electric field (symmetric under $i \longleftrightarrow j$) and $\Tilde{B}_{ij}$ is the rank-2 tensor magnetic field.

In terms of these dual gauge fields, the Lagrangian of the theory can be written only in terms of the electric and magnetic fields, including sources, as:
\begin{align}\label{eq:lagrangian}
\mathcal{L}&=\frac{1}{2}\Tilde{C}^{-1}_{ijkl}E_{ij}E_{kl}-\frac{1}{2\rho_d}(B_i)^{2}+\frac{1}{2\kappa}(\Tilde{E}_{kij}-\epsilon_{ik}\epsilon_{jl}\partial_l\phi)^2-\frac{1}{2}(\Tilde{B}_{ij}+\epsilon_{ik}\epsilon_{jl}A_{kl})^2 + \mathcal{L}_{\rm source}\, ,\\ \nonumber
\mathcal{L}_{\rm source}&=\phi\rho-A_{kl}J_{kl}+\Tilde{\phi}_{ij}\Tilde{\rho}_{ij}-\Tilde{A}_{mkl}\Tilde{J}_{mkl}\, ,   
\end{align}
with $\mathcal{L}_{\rm srcs}$ being discussed below. We note that the above Lagrangian is gauge invariant. That is, although the fields $\Tilde{E}_{kij}$ and $\Tilde{B}_{ij}$ are not gauge invariant, the combinations $(\Tilde{E}_{kij}-\epsilon_{ik}\epsilon_{jl}\partial_l\phi)$ and $(\Tilde{B}_{ij}+\epsilon_{ik}\epsilon_{jl}A_{kl})$ are gauge invariant as required for a dual gauge theory of the elasticity theory considered here.

\section{Sources}
As implicitly shown above, we can write the action $S_{\rm srcs}$ in terms of the gauge fields as:
\begin{align}
S_{\rm source}&=\int d^2xdt\left[\pi_i\partial_t u_{i}^{(s)}+L_{ij}\partial_th_{ij}^{(s)}-(u^{(s)}_{ij}+h^{(s)}_{ij})T_{ij}-\partial_kh_{ij}K_{kij}\right]\\ \nonumber
&=\int d^2xdt\left[\phi\rho-A_{kl}J_{kl}+\Tilde{\phi}_{ij}\Tilde{\rho}_{ij}-\Tilde{A}_{mkl}\Tilde{J}_{mkl}\right] \, , 
\end{align}
by defining the sources in the dual gauge theory as
\begin{align}
\rho&=\epsilon_{ik}\epsilon_{jl}\partial_i\partial_j u_{kl}^{(s)}\, ,\qquad J_{ij}=\epsilon_{ik}\epsilon_{jl}(\partial_t\partial_k-\partial_k\partial_t)u_l^{(s)},\, \\
\Tilde{\rho}_{ij}&=\epsilon_{ik}\epsilon_{jl}\epsilon_{mn}\partial_n\partial_mh^{(s)}_{kl}, \, \qquad \Tilde{J}_{mij}=\epsilon_{ik}\epsilon_{jl}(\partial_m\partial_t-\partial_t\partial_m)h_{kl}^{(s)}\, .  
\end{align}
   
\section{Gauge Transformations}
In our theory,  since by dropping the terms containing metric fluctuations one should recover the original fracton elasticity duality gauge transformations associated to the first set of gauge fields (given by an arbitrary function $f$), we choose to retain those as fixed and work on finding those for the second set of gauge fields (given by an arbitrary function $\Tilde{f}_{kl}$). In this sense we have:
\begin{equation*}
\delta L_{ij}=\epsilon_{ik}\epsilon_{jl}[\partial_m(\delta \Tilde{A}_{mkl})+\delta A_{kl}]=\partial_m(\delta \Tilde{A}_{mkl})-\partial_k\partial_l f=0 \Rightarrow \delta \tilde{A}_{mkl}=\delta_{m\left(k\right.}\partial_{\left.l\right)}f+\tilde{F}_{mkl}.
\end{equation*}
$\tilde{F}_{mkl}$ is a function that is symmetric in the last two indices and also has the symmetry $\partial_m \Tilde{F}_{mkl}=0$. For reasons that will become obvious in a bit we choose to set it equal to
\begin{equation*}
\Tilde{F}_{mkl}=\epsilon_{mn}\partial_n\Tilde{f}_{kl}   . 
\end{equation*}
In addition
\begin{equation*}
\begin{split}
\delta K_{mij}=\epsilon_{ik}\epsilon_{jl}[\partial_t(\delta \Tilde{A}_{mkl})-\delta_{m\left(k\right.}\partial_{\left.l\right)}(\delta \phi)+\epsilon_{mn}\partial_n(\delta \Tilde{\phi}_{kl})]=\hspace{2cm}\\
=\partial_t\left[\delta_{m\left(k\right.}\partial_{\left.l\right)}f+\epsilon_{mn}\partial_n\Tilde{f}_{kl}\right]-\delta_{m\left(k\right.}\partial_{\left.l\right)}\partial_tf+\epsilon_{mn}\partial_n(\delta \Tilde{\phi}_{kl})=\partial_t\partial_n \Tilde{f}_{kl}+\partial_n(\delta \Tilde{\phi}_{kl})=0
\end{split}
\end{equation*}
where the last equality is fulfilled requiring that $\delta \Tilde{\phi}_{kl}=-\partial_t\Tilde{f}_{kl}$. Finally we have,
\begin{align}
     \delta\phi &=\partial_t f\, , \qquad \delta A_{ij}=-\partial_i\partial_jf, \\
     \delta \Tilde{\phi}_{ij}&=-\partial_t\Tilde{f}_{ij},\, \qquad  \delta \Tilde{A}_{mij}=\delta_{m\left(i\right.}\partial_{\left.j\right)}f+\epsilon_{mn}\partial_n\Tilde{f}_{ij} \, .
\end{align}
    
where again we note that $\Tilde{f}_{ij}=\Tilde{f}_{ji}$.

\section{Continuity equations}
The continuity equations for our theory can be derived by
\begin{equation*}
\begin{split}
\delta S_{\rm source}=\int d^2xdt\left[\rho(\delta \phi)-J_{ij}(\delta A_{ij})+\Tilde{\rho}_{ij}(\delta\Tilde{\phi}_{ij})-\Tilde{J}_{mij}(\delta\Tilde{A}_{mij})\right]=\hspace{0.8cm}\\
=\int d^2xdt\left[\rho(\partial_t f)+J_{ij}(\partial_i\partial_j f)-\Tilde{\rho}_{ij}(\partial_t\Tilde{f}_{ij})-\Tilde{J}_{mij}[\delta_{m\left(i\right.}\partial_{\left.j\right)}f+\epsilon_{mn}\partial_n\Tilde{f}_{ij}]\right]=\\
=\int d^2xdt\left[\left(-\partial_t\rho+\partial_i\partial_jJ_{ij}+\partial_j\Tilde{J}_{iij}\right)f+\left(\partial_t\Tilde{\rho}_{ij}-\epsilon_{nm}\partial_n\Tilde{J}_{mij}\right)\Tilde{f}_{ij}\right]=0\, ,
\end{split}
\end{equation*}
from which we obtain
\begin{equation}
    \partial_t\rho-\partial_i\partial_jJ_{ij}-\partial_j\Tilde{J}_{iij}=0\, , \qquad \partial_t\Tilde{\rho}_{ij}-\epsilon_{nm}\partial_n\Tilde{J}_{mij}=0 \, ,
\end{equation}
where the signs are conventional. They are a consequence of how the source terms are introduced in the action.

\section{Equations of motion - Maxwell equations for the symmetric tensor gauge fields}
We can now use the Lagrangian in Eq. \eqref{eq:lagrangian} to derive the Maxwell equations of our theory. 
The two sets of Maxwell equations are as follows:
\vspace{0.25 cm}

\underline{1st\hspace{1mm}set}
\begin{equation*}
 \Tilde{C}_{ijkl}^{-1}\partial _i\partial_jE_{kl}-\left(\frac{1}{\kappa}\epsilon_{ik}\epsilon_{jl}\partial_l\Tilde{E}_{kij}-\frac{2}{\kappa}\partial_m^2\phi\right)=-\rho \, ,  \quad {\rm (Gauss\,\, Law).}
\end{equation*}
\begin{equation} \partial_tB_i+\epsilon_{jk}\partial_jE_{ki}=0\,  ,  \qquad   {\rm (Faraday\,\,Law)} .
\end{equation}
\begin{equation*}
\Tilde{C}_{ijkl}^{-1}\partial_tE_{kl}+\epsilon_{\left(ik\right.}\partial_kB_{\left.j\right)}+\left(\epsilon_{ik}\epsilon_{jl}\Tilde{B}_{kl}+A_{ij}\right)=-J_{ij} \, ,  \quad \text{\rm (Amp\`{e}re\,\,Law),}
\end{equation*}
\label{1stMaxwell}

\underline{2nd\hspace{1mm}set}
\begin{equation*}
 \frac{1}{\kappa}\left(\epsilon_{ik}\epsilon_{jl}\epsilon_{mn}\partial_n\Tilde{E}_{mkl}+\epsilon_{ik}\partial_k\partial_j\phi\right)=-\Tilde{\rho}_{ij}, \qquad {\rm (Gauss\,\, Law),}   
\end{equation*}
\begin{equation}
\partial_t\Tilde{B}_{ij}-\partial_m\Tilde{E}_{mij}=0  ,\qquad  {\rm (Faraday\,\, Law),}  
\end{equation}
\begin{equation*}
\frac{1}{\kappa}\left(\epsilon_{ik}\epsilon_{jl}\partial_t\Tilde{E}_{mkl}-\delta_{mi}\partial_t\partial_j\phi\right)-\epsilon_{ik}\epsilon_{jl}\partial_m\Tilde{B}_{kl}-\partial_mA_{ij}=-\Tilde{J}_{mij} , \qquad \text{\rm (Amp\`{e}re\,\, Law).}  
\end{equation*}
In the first set above, setting the terms in the parentheses equal to zero, one may recover  the laws governing the fracton-elasticity duality derived in \cite{pretko_fracton-elasticity_2018}.

\section{System Hamiltonian}
The Lagrangian \eqref{eq:lagrangian} in terms of the gauge fields reads
\begin{equation}
\begin{split}
&\mathcal{L}=\frac{1}{2}\Tilde{C}_{ijkl}^{-1}(\partial_tA_{ij}+\partial_i\partial_j\phi)(\partial_tA_{kl}+\partial_k\partial_l\phi)-\frac{1}{2\rho_d}\partial_mA_{ni}(\partial_mA_{ni}-\partial_nA_{mi})+\\
&+\frac{1}{2\kappa}\left[\epsilon_{ik}\epsilon_{jl}(\partial_t\Tilde{A}_{mkl}+\epsilon_{mn}\partial_n\Tilde{\phi}_{kl})-\epsilon_{im}\epsilon_{jl}\partial_l\phi\right]^2-\frac{1}{2}\left(\epsilon_{ik}\epsilon_{jl}\partial_m\Tilde{A}_{mkl}+\epsilon_{ik}\epsilon_{jl}A_{kl}\right)^2+\\
&+\phi\rho-A_{ij}J_{ij}+\Tilde{\phi}_{ij}\Tilde{\rho}_{ij}-\Tilde{A}_{mkl}\Tilde{J}_{mkl}.
\end{split}
\end{equation}
In order to write the Hamiltonian density we obtain the canonical momenta which are given by:
\begin{equation}
\Pi_{ij}=\left(\frac{\partial \mathcal{L}}{\partial \Dot{A}_{ij}}\right)=\Tilde{C}_{ijkl}^{-1}(\partial_tA_{ij}+\partial_i\partial_j\phi)=-\Tilde{C}_{ijkl}^{-1}E_{kl} ,
\end{equation}
\begin{equation}
\Tilde{\Pi}_{ijk}=\left(\frac{\partial \mathcal{L}}{\partial \Dot{\Tilde{A}}_{ijk}}\right)=\frac{1}{\kappa}\left(\partial_t\Tilde{A}_{ijk}+\epsilon_{im}\partial_m\Tilde{\phi}_{jk}-\delta_{ij}\partial_k\phi\right)    .
\end{equation}
Conjugated momenta for the fields $\phi$ and $\Tilde{\phi}_{ij}$ cannot be defined as the Lagrangian does not depend on their time derivatives. They instead behave as Lagrange multipliers. With this, the Hamiltonian density can be written as,
\begin{equation}
\mathcal{H}=\Pi_{ij}\Dot{A}_{ij}+\Tilde{\Pi}_{ijk}\Dot{\Tilde{A}}_{ijk}-\mathcal{L}.
\end{equation}
Varying $\mathcal{H}$ above and using the conservation equations and the form of the gauge transformations stated above, one obtains the same Gauss law's derived before as
\begin{equation}
\partial_i\partial_j\Pi_{ij}+\partial_j\Tilde{\Pi}_{iij}=-\rho \, , \qquad    \epsilon_{km}\partial_m\Tilde{\Pi}_{kij}=-\Tilde{\rho}_{ij}  \, . 
\end{equation}

\section{Fractonic Behavior}
Following an argument similar to \cite{pretko_crystal--fracton_2019} we first compute the dipole moment associated to the charges in our theory. For the charge $\rho$ associated with the first set of gauge fields we have
\begin{align*}
\begin{split}
 P_a&=\int \rho\,  x_a d^2x=-\epsilon_{im}\epsilon_{jn}\int\left[\Tilde{C}_{ijkl}^{-1}\partial_m\partial_nT_{kl}+\frac{1}{\kappa}\partial_mK_{nij}\right]x_ad^2x\\
&=-\epsilon_{im}\epsilon_{jn}\int\left[\partial_m\left(\Tilde{C}_{ijkl}^{-1}x_a\partial_nT_{kl}\right)-\delta_{ma}\left(\Tilde{C}_{ijkl}^{-1}\partial_nT_{kl}\right)+\frac{1}{\kappa}\partial_m\left(x_aK_{mij}\right)-\frac{1}{\kappa}\delta_{ma}K_{nij}\right]d^2x\, .
\end{split}
\end{align*}
Noting that the total divergences were switched to surface integrals through the Gauss-Ostrogradsky theorem, one may write
\begin{equation}
 P_a-\frac{1}{\kappa}\int\left(K_{aii}-K_{iia}\right)d^2x=\text{const}\, .    
\end{equation}
 For the tensor charge $\Tilde{\rho}_{kl}$ associated to the second set of gauge fields, we define the dipole moment as,
\begin{equation}
\begin{split}
\Tilde{P}_l&=\int \epsilon_{il}\Tilde{\rho}_{ij}x_jd^2x=\frac{1}{\kappa}\int \epsilon_{il}\epsilon_{im}\epsilon_{jn}\epsilon_{kp}\partial_pK_{kmn}x_jd^2x=\frac{1}{\kappa}\int \epsilon_{jn}\epsilon_{kp}\partial_p K_{kln}x_jd^2x=\\
&=\frac{\epsilon_{jn}\epsilon_{kp}}{\kappa}\int\left[\partial_p\left(K_{kln}x_j\right)-\delta_{jp}K_{kln}\right]d^2x\Rightarrow \Tilde{P}_l-\frac{1}{\kappa}\int K_{iil}d^2x=\text{const}.
\end{split}    
\end{equation}
Therefore, for the total dipole moment of our theory we have
\begin{equation}\label{eq:dipole_cons}
P_a+\Tilde{P}_a=\frac{1}{\kappa}\int K_{aii}d^2x+\text{const}  \, . 
\end{equation}
Interestingly, when $K_{aii}=0$, the total dipole moment is conserved. This is analogous to the case of  Cosserat elasticity in which one also gets two independent dipole moments associated  with  two different kind of defects \cite{gromov_duality_2020}. This shows that disclinations can have fractonic behavior. To this end, we follow \cite{pretko_fracton-elasticity_2018} and we express the right-hand side of \eqref{eq:dipole_cons} in terms of elastic variables as,
\begin{equation}
\frac{1}{\kappa}K_{aii}=\partial_ah_{ii}=\partial_a(\Bar{h}_{ii}+h^{(s)}_{ii})=0   \, . 
\end{equation}
The smooth part denoted by the bar represents regular volume changes attributed to the regular changes in the metric. These, as noted in \cite{pretko_fracton-elasticity_2018} regarding  creation of vacancies in the crystal can sensibly  be set to zero i.e. $\partial_a\Bar{h}_{ii}=0$. Therefore, the singular part now represents volume changes depending on the disclinational defects present on the metric. Should that term be zero, i.e.
\begin{equation}
    \partial_ah_{ii}^{(s)}=0\, .
\end{equation} 
then indeed the disclinations behave like fractons as the total dipole moment of the system is conserved. Assuming this condition is fulfilled, this immediately implies 
\begin{align}
    \Tilde{\rho}_{ii}&=\epsilon_{mn}\partial_n\partial_mh^{(s)}_{ii}=0 ,\, \\ \nonumber 
    \Tilde{J}_{mii}&=\partial_m\partial_th^{(s)}_{ii} \, . 
\end{align}
Now, in order to analyze the dislocations, we have to look into the quadrupole moment. Regarding the scalar charge $\rho$ we have
\begin{equation*}
\begin{split}
Q_{ii}&=\int \rho\, x_i^2d^2x=-\epsilon_{im}\epsilon_{jn}\int\left(\Tilde{C}_{ijkl}^{-1}\partial_m\partial_nT_{kl}+\frac{1}{\kappa}\partial_mK_{nij}\right)x_i^2d^2x\\
&=-\epsilon_{im}\epsilon_{jn}\int\left[\Tilde{C}_{ijkl}^{-1}\partial_m(\partial_nT_{kl}x^2)-\Tilde{C}_{ijkl}^{-1}(\partial_nT_{kl})(\partial_mx^2)+\frac{1}{\kappa}\partial_m(K_{nij}x^2)-\frac{2}{\kappa}K_{nij}x_m\right]d^2x\\
&=\epsilon_{im}\epsilon_{jn}\int\left(\Tilde{C}_{ijkl}^{-1}\partial_n(2T_{kl}x_m)+2\Tilde{C}_{ijkl}^{-1}T_{kl}\delta_{mn}-\frac{2}{\kappa}K_{nij}x_m\right)d^2x+\text{const}\, ,
\end{split}
\end{equation*}
which can be compactly written as
\begin{equation}
Q_{ii}=2\int\left[\Tilde{C}_{iikl}^{-1}T_{kl}-\frac{1}{\kappa}\left(K_{kkl}x_l-K_{kll}x_k\right)\right]d^2x+\text{const}.    
\end{equation}\\
On the other hand, referring to the tensor charge $\Tilde{\rho}_{ij}$ we have
\begin{equation*}
\begin{split}
\Tilde{Q}_{ml}&=\int \epsilon_{il}\Tilde{\rho}_{ij}x_j x_md^2x=\frac{1}{\kappa}\int \epsilon_{jn}\epsilon_{kp}\partial_pK_{kln}x_jx_md^2x=\frac{\epsilon_{jn}\epsilon_{kl}}{\kappa}\int \partial_pK_{kln}x_jx_md^2x=\\
&=\frac{\epsilon_{jn}\epsilon_{kl}}{\kappa}\int\left[\partial_p(K_{kln}x_jx_m)-K_{kln}\partial_p(x_jx_m)\right]d^2x=-\frac{\epsilon_{jn}\epsilon_{kp}}{\kappa}\int K_{kln}(\delta_{pj}x_m+x_j\delta_{pm})d^2x=\\
&=\frac{1}{\kappa}\int \left(K_{kkl}x_m-\epsilon_{jn}\epsilon_{km}K_{kln}x_j\right)d^2x+\text{const}\,
\end{split}
\end{equation*}
which finally gives
\begin{equation}
 \Tilde{Q}_{ml}=\frac{1}{\kappa}\int\left(2K_{kkl}x_m-K_{klm}x_k\right)d^2x+\text{const}.  \end{equation}
With this, by taking the trace on $\Tilde{Q}_{ml}$ one may write
\begin{equation}\label{eq:quadrupole}
Q_{ii}+\Tilde{Q}_{ii}=\int \left[2\Tilde{C}_{iikl}^{-1}T_{kl}+\frac{1}{\kappa}K_{aii}x_a\right]d^2x+\text{const}\, .
\end{equation}
With this, requiring that the dipole moment is conserved implies that the second term in the right-hand  side in \eqref{eq:quadrupole} is zero. The first term being zero  is exactly the regular constraint imposed in \cite{pretko_fracton-elasticity_2018} along with a term originating from the metric that has a similar contribution (the stress-tensor $T_{ij}$ is dual to $w_{ij}$).

We have for each case separately the following:

\underline{1st case}: The trace of both the strain tensor and the metric tensor are equal to zero. In this simple case which proves to be the analog of Pretko's theory in curved space, the densities and currents simplify to:
\begin{equation}
u_{ii}^{(s)}=h_{ii}^{(s)}=0 ,\hspace{2cm} \Tilde{\rho}_{ii}=0 ,\hspace{2cm}J_{ii}=0 ,\hspace{2cm} \Tilde{J}_{mii}=0 ,   
\end{equation}
\begin{equation}
\rho=-\partial_i\partial_ju_{ij}^{(s)} ,\hspace{2cm} \Tilde{\rho}_{ij}=\epsilon_{nm}\partial_n\partial_mh_{ij}^{(s)},
\end{equation}
\begin{equation}
J_{ij}=\partial_i\partial_tu_j^{(s)}-\partial_k\partial_tu_{k}^{(s)}\delta_{ij}, \hspace{2cm} \Tilde{J}_{mij}=(\partial_t\partial_m-\partial_m\partial_t)h_{ij}^{(s)}.
\end{equation}
The defects coming from the displacement field and the metric both contribute to the volume changes but without a relation between their currents.\\

\underline{2nd case}: Here we require that the generalized strain tensor $w_{ij}$ be equal to zero in order to cancel the contribution for the volume changes in the first term of Eq. \eqref{eq:quadrupole}. In total we have the conditions:
\begin{equation}
u_{ii}^{(s)}=-h_{ii}^{(s)} ,\hspace{2cm} \partial_a u_{ii}^{(s)}= \partial_a h_{ii}^{(s)}=0, \hspace{2cm}  \Tilde{\rho}_{ii}=0 ,\hspace{2cm} 
\end{equation}
\begin{equation}
\rho=\partial^2u_{kk}^{(s)}-\partial_i\partial_ju_{ij}^{(s)} ,\hspace{2cm}\Tilde{\rho}_{ij}=\epsilon_{nm}\partial_n\partial_mh_{ij}^{(s)}  ,   
\end{equation}
\begin{equation}
J_{ij}=(\partial_i\partial_t-\partial_t\partial_i)u_j^{(s)}-\partial_k\partial_tu_k^{(s)}\delta_{ij} , \hspace{2cm}  \Tilde{J}_{mij}=-\partial_m\partial_t(u_{kk}^{(s)}\delta_{ij}+h_{ij}^{(s)}),
\end{equation}
\begin{equation}
\Tilde{J}_{mii}=-\partial_m\partial_t u_{ii}^{(s)} ,\hspace{2cm}  J_{ii}=\partial_t\partial_iu_i^{(s)}-\partial_i\partial_tu_i^{(s)} .
\end{equation}
By combining the two relations above we get:
\begin{equation}
\Tilde{J}_{mii}=-(\partial_mJ_{ii}+\partial_m\partial_i\partial_tu_i^{(s)})=-(\partial_mJ_{ii}+\partial_m\partial_i\pi_i^{(s)})\, .    
\end{equation}
The Amp\`{e}re Law in the first set of Maxwell equations, can be rewritten to take the form of a continuity equation if someone reverts back to elastic variables. We have:
\begin{equation}
\partial_tn_d+\partial_i\pi_i^{(s)}+\partial_t\Tilde{\pi}^{(s)}=-J_{ii}\, ,    
\end{equation}
where $\Tilde{\pi}^{(s)}=L_{ii}^{(s)}=\partial_th_{ii}^{(s)}$ and $n_d$ the vacancy-interstitial density coming from the displacement field. This relation is again the curved background counterpart of the one Pretko derived regarding the continuity equation for the defects. We now combine the last two to arrive at:
\begin{equation}
\Tilde{J}_{mii}=\partial_m\partial_t\left(n_d+\Tilde{\pi}^{(s)}\right) .   
\end{equation}
The singular part of the trace of the metric dual tensor $L_{ij}$ effectively works as the vacancy-interstitial density along with $n_d$ for the metric defect current.

\end{document}